\newcommand{\za}{\alpha}

\newcommand{\zs}{\sigma}

\newcommand{\zl}{\lambda}

\newcommand{\zn}{\nu}

\newcommand{\zr}{\rho}
\newcommand{\zt}{\tau}
\newcommand{\zw}{\omega}

\newcommand{\zF}{\Phi}
\newcommand{\zG}{\Gamma}
\newcommand{\zL}{\Lambda}
\newcommand{\zN}{I\hskip-3.4pt N}
\newcommand{\zR}{I\hskip-3.4pt R}

\newcommand{\zW}{\Omega}

\newcommand{\doublespace}
  {\addtolength{\baselineskip}{.5\baselineskip}}

\newcommand{\currentspace}{} 

\documentstyle[11pt]{article}

\newlength{\fullboxwidth}
\begin{document}

\title{Orbital measures in non-equilibrium statistical mechanics: 
the Onsager relations}
\author{Short title: Orbital Measures in Statistical Mechanics \\ \\ \\
L. Rondoni \\ Dipartimento di Matematica, Politecnico di Torino\\
Corso Duca degli Abruzzi 24, I-10129 Torino, Italy \\ \\
E.G.D. Cohen \\
Rockefeller University, New York, New York 10021 - U.S.A.\\
\\
e-mail:  rondoni @ polito.it \\ \\ \\
Physics abstracts numbers: 05.45.+b, 05.60.+w, 05.70.Ln \\
AMS classification unmbers: 82C05, 70F25, 82C70}

\maketitle

\pagestyle{myheadings}
\markright{Orbital Measures in Statistical Mechanics}

\currentspace{\doublespace} 

\noindent
{\Large \bf Abstract}
\vskip 10pt \noindent
We assume that the properties of nonequilibrium stationary states
of systems of particles can be expressed in terms of weighted
orbital measures, i.e. through periodic orbit expansions. 
This allows us to derive the Onsager relations
for systems of $N$ particles
subject to a Gaussian thermostat, under
the assumption that the entropy production rate is equal to
the phase space contraction rate. 
Moreover, this also allows us 
to prove that the relevant transport coefficients are not negative.
In the appendix we give an argument for the proper way of treating
grazing collisions, a source of possible singularities in the dynamics.

\bigskip

\section{Introduction}

Very recently a number of proofs have been presented of the
Onsager reciprocal relations (OR) in non-equilibrium stationary states
based on dynamical systems theory \cite{G1,G2,GR}. There,
the earlier assumption of Refs.\cite{GC1,GC2}, i.e. that the dynamics of 
systems of particles could be regarded as ``Anosov'' for practical
purposes,\footnote{For completeness, the main assumption of
Ref.\cite{GC2}, p.935 states that:

\noindent
{\em A reversible, many particle system in 
a stationary state, can be regarded as a transitive 
Anosov system, for the purpose of computing its macroscopic properties}. 

\noindent
In other words, if the (possible) deviations of the dynamics 
from the Anosov conditions bear no consequence for the calculations of
the quantities of physical interest, they can be considered as 
irrelevant. In that case, the system will be called ``Anosov-like''.}
was used to
characterize the relevant stationary distributions. Consequently, such
distributions were 
conjectured to be of the kind first investigated by Ya. Sinai, D. Ruelle 
and R. Bowen (SRB distributions) \cite{SRB}. This
assumption makes possible a generalization
to nonequilibrium stationary states of the classical equilibrium
ensembles for the probability to find a system in a certain phase
in phase space. In particular, this can be done by attributing to
the absolute values of the Jacobian
of the dynamics restricted to the unstable manifolds, a meaning
connected to that of the Boltzmann factor of the 
canonical ensemble (see subsection 2.2 below, and see
Ref.\cite{MRC} for numerical experiments meant to justify
the connection). Use of such techniques lead to
a fluctuation theorem $-$first investigated in Ref.\cite{ECM2,GC1,GC2}
and further confirmed by
a variety of computer experiments \cite{BGG,LLP}$-$ and 
to the above mentioned proofs of the OR.
 
There are two representations of the SRB distribution which have been
used in practice so far: the Markov partition method of Ref.\cite{GC2},
and the periodic orbit (or cycle) expansions (POE) (see, e.g. 
Refs.\cite{RU,GA,PA,AAC}) used, for instance, in Refs.\cite{CGS,V,MR,LNRM}.
Here, we present an approach to the proof of the OR 
based on the assumption that the stationary
distribution can be expressed in terms of POE, and on analyticity
conditions on the properties of the relevant orbits. The point is not to
improve on the quite general proof of Ref.\cite{GR}, but to show that a
different approach with different assumptions can lead to
similar results, thus shedding new light on the precise connection
between the OR in statistical mechanics 
and the dynamics of mechanical systems.
Periodic orbit expansions are also used, here, for a proof  
that the transport coefficients of our systems
are nonnegative, following
in part ideas developed previously in Ref.\cite{RMlarge} (see also
the paper by D. Ruelle, Ref.\cite{RU0}) based on the equality of phase
space contraction rate and entropy production rate.

The possibility of describing the stationary states of systems of
particles in terms of POE is implied by the assumptions of 
Refs.\cite{GC1,GC2},\footnote{If the system under study
is really Anosov, weighted averages of 
smooth functions evaluated over periodic orbits do converge to
the averages computed with the SRB distribution (see e.g.
Refs.\cite{RU,GA,PA,AAC}), despite the fact that periodic orbits 
cover but a set of zero (Lebesgue and natural) measure. If the system
is only ``Anosov-like'', POE should work as well, at least
for the calculations of the physically interesting quantites. 
However, in this paper we will assume the validity of POE only
for the calculation of the average currents, in order to 
limit the assumptions as much as possible.}
but the actual range of applicability
could be different (see next section). The importance of this 
approach lies
in the fact that, although the practical implementation of the POE
for a system of many particles is exceedingly difficult
(practically impossible with present day technology), its mathematical
expression can be used for formal proofs of properties of real systems.
Moreover, the validity of given results obtained in terms of POE can be 
tested directly on simple models of statistical mechanical
interest, by means of modern computers. In fact, POE have
been succesfully applied in numerical simulations of
one particle systems, such as various (periodic)
Lorentz gas models, see e.g. Refs.\cite{MRC,CGS,V,MR,LNRM}.
In this respect, it is also important to note 
that the efficiency of this kind of 
numerical studies can be greatly enhanced, as recently shown in
Ref.\cite{DM}, thus making the numerical implemetation of POE
conceivable for more complex systems.

\section{Properties of transport coefficients from POE} 

\subsection{Preliminaries}

In this paper, we concentrate on a $d$-dimensional system of $N$ 
interacting particles, subject to external fields of
force (whose strength is gauged by $\zn$ parameters),  
and subject to a Gaussian thermostat \cite{Ho}. Moreover, in order to
make use of the techniques of dynamical system theory, we
impose periodic boundary conditions on our models. 
In particular, the equations of motion we consider
take the form
\begin{equation}
\dot{\bf q}_i = \frac{{\bf p}_i}{m_i} ~,  \quad
\dot{\bf p}_i = {\bf \zF}_i({\bf q}) + {\bf F}_i({\bf G}) - 
\za({\bf p},{\bf q}) {\bf p}_i ~, 
\quad i = 1, ..., N ~,
\label{eqsmot}
\end{equation}
where $\za({\bf p},{\bf q})$ characterizes the Gaussian thermostat and
is determined by the constraint one wants to implement
(e.g. fixing the value of a special dynamical quantity). Also,
${\bf q}_i$, ${\bf p}_i$ and $m_i$ are 
respectively the coordinates,
the momentum and the mass of the $i$th particle, with
$({\bf p},{\bf q}) = (({\bf p}_i)_{i=1}^N,({\bf q}_i)_{i=1}^N)$; 
${\bf \zF}_i$ is the
force on particle $i$ due to the interactions with the other
particles; and ${\bf F}_i$ is the action of the external fields
on the $i$th particle, which depends on the values of the parameters
${\bf G} = (G_1, ..., G_\zn)$. The purpose of the Gaussian thermostat 
is to make such an $N$-particle system reach a stationary state when 
the fields are ``switched on'', starting from a given initial 
distribution in phase space. We assume that when ${\bf G}$ vanishes,  
the external forces ${\bf F}_i$ and the thermostat
coupling $\za$ also vanish, making the
stationary state an equilibrium state.

The particular constraint we impose on our models
is that the total energy remains fixed in
time (i.e. we use an iso-energetic thermostat). This yields
\begin{equation}
\za(\zG;{\bf G}) = \frac{1}{2 K} \sum_{i=1}^N \frac{{\bf p}_i \cdot 
{\bf F}_i({\bf G})}{m_i} ~; \quad \quad K = \sum_{i=1}^N 
\frac{{\bf p}_i^2}{2 m_i}
\label{alfa}
\end{equation}
where $\zG=({\bf p},{\bf q})$ stands for a generic point in the phase 
space, and $K$ is the total kinetic energy of the system.\footnote{In 
case of hard core particles, the iso-energetic
constraint coincides with the iso-kinetic one, i.e. with the 
constraint of constant kinetic energy.}
Once the stationary state has been achieved, the periodic boundary
conditions\footnote{Periodic boundaries may look as a rather restricitve
condition. However, one should keep in mind that $\zW$ can be arbitrarily
large and complex, as long as it may be used to periodically tile the
infinite phase space.}
imply that there is an elementary cell (EC), $\zW$ say, 
whose replicas cover the whole $(2dN)$-dimensional phase space,
and there is a time evolution on $\zW$, denoted by
$S_t$, which represents the dynamics of the $N$ particles in $\zW$.
One may also consider the dynamics with respect to
a given Poincar\'{e} section, ${\cal P}$ say, and 
the timing events for the definition of ${\cal P}$ may be chosen to 
be the collisions between particles. When the interacting
potentials are soft, this can be done by calling a collision the 
event in which two or more particles come within a certain distance 
of each other \cite{GC2}. 

By identifying the phase space contraction rate for 
the dynamics $-$the divergence of the right hand side 
of Eqs.(\ref{eqsmot})$-$
with the (microscopic) entropy production rate, 
we define the $\zn$ (microscopic) {\bf currents} as
\begin{equation}
{\bf J}(\Gamma;{\bf G}) = \nabla_{\bf G} \za(\zG;{\bf G})
\equiv (J_1, ..., J_\zn)(\Gamma;{\bf G}) ~ .
\label{currentdef}
\end{equation}
This definition of current was first adopted in Ref.\cite{G2},
and also used in Refs.\cite{GR,RMlarge}.

In order to develop our analysis in terms of POE, we need that
Unstable Periodic Orbits (UPOs) be densely embedded in the support of 
the natural measure of the system and, if they are,\footnote{Typical 
dynamical systems whose UPOs are dense in the support of 
the natural measure are those verifying the axiom-A conditions. 
Indeed, for such systems
the density condition is part of their very definition.} we must 
characterize their properties. In the first place, UPOs are orbits
with at least one positive Lyapunov exponent, and it is convenient to
collect them in sets of orbits with a fixed number of collisions. 
For instance, denote by
$P_n(G)$ the set of orbits within which $n$ collisions occur. 
Then, we let $\zw(G)$ indicate a generic 
element of $P_n(G)$, and $\tau_{\zw(G)}$ its period. 
Given any orbit $\zw(G)$, with Lyapunov exponents $\zl_{\zw(G),1} 
\ge \zl_{ \zw(G),2} \ge ...
\ge \zl_{ \zw(G),2dN}$, we consider the following quantity:
\begin{equation}
\Lambda_{\zw(G),u}^{-1} = \exp \left( - \tau_{\zw(G)} 
\sum\mbox{$^\prime$} ~ \lambda_{\zw(G),j} \right) \, ,
\label{weidef}
\end{equation}
where $\lambda_{\zw(G),j}$ is the $j$-th Lyapunov
exponent of orbit $\zw(G)$, and the prime on the symbol $\sum$
indicates that the sum runs
over the {\it positive} exponents only. The subscript
$u$ on the left hand side of Eq.(\ref{weidef}) says that
$\Lambda_{\zw(G),u}$ is computed restricting the dynamics
to the unstable manifold. Notice that  
$\Lambda_{\zw(G),u}$ is a
measure of the instability in the phase space of $\zw(G)$, and it
constitutes the basic ingredient
in any POE, i.e. it is part of the
(unnormalized) weight attributed to $\zw(G)$. In particular,
the less unstable the UPO, the larger its weight, as explained in
Section 2.2.

For the study of transport in our particle systems, it is interesting
to consider the total displacement of the single particles 
during a whole period of a UPO. In this respect, given $\zw(G)$,
it is important to recall that the displacement of the $j$th particle
at the end of one period of $\zw(G)$, $\triangle {\bf q}_{j,\zw(G)}$, 
does not necessarily
vanish. When it does not, also the average current associated with 
$\zw(G)$ may not vanish. Here the average current is the integral of 
$\nabla_{\bf G} \za$ over the UPO (that is, 
the cumulative current ${\bf I}_\zw$ defined below in 
Eq.(\ref{cumcur})), divided by the period $\zt_{\zw(G)}$.
For this reason, it makes sense to talk of currents
associated with a UPO although, at first sight,
it may look strange. Indeed, the periodic boundary conditions
impose translation invariance symmetry along the directions of
the lattice vectors in the stationary state.
Hence, to follow a trajectory which never returns to the cell
where it started from, is equivalent to following its translated
image in the EC. 
\begin{figure}
\begin{center}
\setlength{\unitlength}{0.00083300in}%
\begingroup\makeatletter\ifx\SetFigFont\undefined
\def\x#1#2#3#4#5#6#7\relax{\def\x{#1#2#3#4#5#6}}%
\expandafter\x\fmtname xxxxxx\relax \def\y{splain}%
\ifx\x\y   
\gdef\SetFigFont#1#2#3{%
  \ifnum #1<17\tiny\else \ifnum #1<20\small\else
  \ifnum #1<24\normalsize\else \ifnum #1<29\large\else
  \ifnum #1<34\Large\else \ifnum #1<41\LARGE\else
     \huge\fi\fi\fi\fi\fi\fi
  \csname #3\endcsname}%
\else
\gdef\SetFigFont#1#2#3{\begingroup
  \count@#1\relax \ifnum 25<\count@\count@25\fi
  \def\x{\endgroup\@setsize\SetFigFont{#2pt}}%
  \expandafter\x
    \csname \romannumeral\the\count@ pt\expandafter\endcsname
    \csname @\romannumeral\the\count@ pt\endcsname
  \csname #3\endcsname}%
\fi
\fi\endgroup
\begin{picture}(6324,1926)(1714,-5050)
\thicklines
\put(2476,-4186){\circle*{336}}
\put(7020,-3511){\circle*{212}}
\put(7351,-4111){\circle*{336}}
\put(1726,-4561){\framebox(1500,1425){}}
\put(3226,-4561){\framebox(1500,1425){}}
\put(2476,-3511){\vector( 1, 0){1275}}
\put(2326,-3511){\circle*{212}}
\put(2626,-4186){\vector( 1, 0){1275}}
\put(7276,-5011){\makebox(0,0)[lb]{\smash{\SetFigFont{12}{16.8}{rm}(b)}}}
\put(6526,-3511){\vector( 1, 0){225}}
\put(6526,-4561){\framebox(1500,1425){}}
\put(7201,-3511){\line( 1, 0){825}}
\put(6526,-4111){\vector( 1, 0){525}}
\put(7651,-4111){\line( 1, 0){375}}
\put(3076,-5011){\makebox(0,0)[lb]{\smash{\SetFigFont{12}{14.4}{rm}(a)}}}
\end{picture}
\caption{The equivalence between the dynamics in the EC and 
those in the full phase space. a) Trajectories of two black round
particles in the infinite phase space where they
move never returning to the same point. Only two adjacent cells,
out of the infinitely many which tile the phase space, are drawn.
b) Representation of the situation (a) in the EC, where the orbits appear
to be periodic. To average any quantity over the trajectories in (a)
is the same as to average that quantity over the trajectories in (b),
as long as the actual diplacements are accounted for.}
\end{center}
\end{figure}
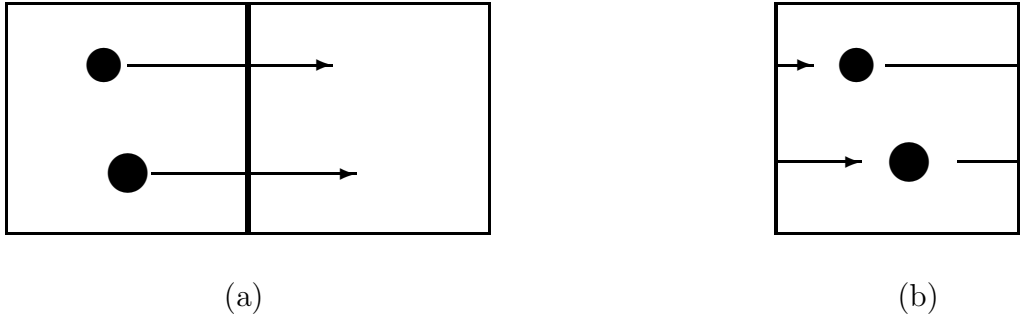
See, e.g., Fig.1: the trajectories of the round black  
particles in the full phase space do not close. However, 
they can be followed in
the EC, where they appear to close producing a periodic orbit.
Clearly, to average quantities over the open paths $-$the real paths
in phase space$-$ is the
same as averaging them over the images of such paths in the EC,
as long as one keeps track of the actual distance covered
during the motion. 
In this sense, when
we speak of periodic orbits, we mean all those that are such in the EC.
In particular, if the particles in Fig.1 carry the same charge $c$, 
the side of the EC is ${\ell}$, and the period of the orbit is $\zt$, 
the relevant current is $2 c {\ell} / \zt$.
We remark again that 
it is the presence of the periodic boundary conditions
which, {\em in the stationary state}, allows us to map the infinite system
into a finite one: the EC. Away from the stationary state, or in the
case that we do not have periodic boundary conditions (like in the random 
Lorentz gas), the techniques of dynamical system theory fail to apply
in this simple manner.

\subsection{Conditions} 

The present paper rests
on Assumptions 1. and 2. below which, 
similarly to those of Refs.\cite{GC1,GC2}, are empirically motivated. 
In particular, our assumptions were 
inspired by the numerical results presented in 
Refs.\cite{MRC,CGS,V,MR,LNRM} and, in a broader sense, by those
of Ref.\cite{ECM2}.

Let us introduce the quantity ${\bf I}_{\zw(G)}$, the {\bf cumulative
current} associated with $\zw(G) \in P_n(G)$:
\begin{equation}
{\bf I}_{\zw(G)} = \int_{\zw(G)} {\bf J}(S_t \zG_{\zw(G)}) ~ dt ~,
\label{cumcur}
\end{equation}
where the integral is carried over one period of $\zw(G)$, and
$S_t \zG_{\zw(G)}$ is the point representing the state of the 
system at time $t$, if it was $\zG_{\zw(G)} \in \zw(G)$ at $t=0$.
Then, our first assumption can be formulated as follows.
\newtheorem{assumption}{Assumption}
\begin{assumption} {\bf (Existence and representation of stationary
states).}
The dynamical system $(\Omega,S_t)$ describing a reversible many-particle
system obeys the ``Extended zero-th law''
of Refs.\cite{GC1,GC2}, at least for what concerns the currents.
Then, letting $\mu_G$ be the corresponding (SRB) stationary distribution,
we can write
\begin{equation}
\langle {\bf J} \rangle_{_{\zW}}({\bf G}) \equiv
\lim_{T\rightarrow\infty} \frac{1}{T} \int_0^T {\bf J}(S_t \zG) ~d t
= \int_\zW {\bf J}(\zG) ~ d\mu_G(\zG) ~, 
\end{equation}
for (Lebesgue) almost all $\zG \in \zW$.
Also, the phase space average of ${\bf J}$ with distribution 
$\mu_G$ can be given in terms of the following POE:
\begin{equation}
\langle {\bf J} \rangle_{_{\Omega}}({\bf G}) = 
\lim_{n \rightarrow \infty}
\frac{\sum_{\zw \in P_n(G)} \tau_\zw \Lambda_{\zw,u}^{-1} 
{\bf I}_\zw / \tau_\zw}{
\sum_{\zw \in P_n(G)} \tau_\zw \Lambda_{\zw,u}^{-1}} =
\lim_{n \rightarrow \infty}
\frac{\sum_{\zw \in P_n(G)} \Lambda_{\zw,u}^{-1} 
{\bf I}_\zw}{
\sum_{\zw \in P_n(G)} \tau_\zw \Lambda_{\zw,u}^{-1}} \, .
\label{POEJ}
\end{equation} 
\end{assumption}
Equation (\ref{POEJ}) expresses the current as a limit of weighted averages
of orbital average currents, where the weights have the suggestive form
\begin{equation}
\zt_\zw \zL_{\zw,u}^{-1} =  \mbox{(time spent in $\zw$) $\times$ 
(inverse of instability of $\zw$)} ~,
\end{equation}
apart from a normalization factor. This weight associates the points of 
a given region of phase space with a probability and is
larger for longer UPOs, while it is
smaller for more unstable UPOs. It is in this sense
that $\zt_\zw \zL_{\zw,u}^{-1}$ plays a role similar to that of the 
Boltzmann factor in the canonical ensemble.

Let us denote by $\langle {\bf J} \rangle^{(n)}$ the ``$n$-th order UPO 
approximation'' of the current, i.e.:
\begin{equation}
\langle {\bf J} \rangle^{(n)}({\bf G}) = 
\frac{\sum_{\zw \in P_n(G)} \Lambda_{\zw,u}^{-1} {\bf I}_{\zw}}
{\sum_{\zw \in P_n(G)} \tau_\zw \Lambda_{\zw,u}^{-1}} ~ ,
\label{Jk}
\end{equation}
where all properties of a given UPO, $\zw$ say, depend on ${\bf G}$, 
because $\zw$ is taken from the set $P_n(G)$. 
\begin{assumption} {\bf (Properties near equilibrium).} There is a
positive constant $\zr$ such that for $|{\bf G}|$ (the modulus of
{\bf G}) smaller than $\zr$ we have: \begin{enumerate}
\item[{\bf a.}] the
period $\tau_{\zw(G)}$, the stability weight $\Lambda_{\zw(G),u}^{-1}$,
and the thermostat coupling $\za(S_t \zG_\zw;{\bf G})$, are ``left'' 
and ``right'' analytic\footnote{Given a function $f : \zR \rightarrow \zR$, 
and a constant $\zr > 0$,
we say that $f$ is {\em left analytic} in $(-\zr,0]$ if there are 
constants $f_0$ and $r^-$ in $\zR$ such that
$f(x) = f_0 +  r^- x + o(x^2)$, for all $x \in (-\zr,0]$.
Similarly, we say that $f$ is {\em right analytic} in $[0,\zr)$ if there are
constants $f_0$ and $r^+$ in $\zR$ such that
$f(x) = f_0 +  r^+ x + o(x^2)$, for all $x \in [0,\zr)$.
Thus, a function which is analytic in a neighborhood of zero
is left and right analytic, and has $r^- = r^+$.
However, $f$ could be left and right analytic without its derivatives
being defined at $x=0$. 
Here $o(x^2)$ is to be understood as a term of second order in $x$, 
negligible with respect to the other terms for all $x$ in the relevant
intervals.}
with respect to all the components of {\bf G}, for
all $\zw(G) \in P_n(G)$, all $n \in \zN$, and all $t \ge 0$. 
\item[{\bf b.}] $\{ \partial_{G_i} \langle {\bf J} \rangle^{(n)}({\bf G}) \}$,
converges uniformly to a limit as $n \rightarrow \infty$, 
for $i = 1, ..., \zn$.
\item[{\bf c.}] The support of $\mu_G$ is dense in $\zW$.
\end{enumerate}
\end{assumption}
In particular, this assumption implies the existence of
constants ${\bf a}_\zw^+, {\bf a}_\zw^-, {\bf b}_\zw^+, {\bf b}_\zw^-,
{\bf c}_\zw^+$ and $ {\bf c}_\zw^-$ in $\zR^\zn$ such that
\begin{eqnarray}
& &\tau_{\zw(G)} = \tau_{\zw(0)} + {\bf a}_{\zw} \cdot {\bf G} + 
o(G^2) ~, \label{tayT} \\
& &\Lambda_{\zw(G),u}^{-1} = \Lambda_{\zw(0),u}^{-1} + {\bf b}_{\zw} 
\cdot {\bf G} + o(G^2) \label{tayL} ~, \\
& &\za(S_t \zG_{\zw(G)}) = {\bf c}_{\zw(0)}(S_t \zG_{\zw(0)}) 
\cdot {\bf G} + o(G^2) ~, \label{tayX}
\end{eqnarray}
where the $i$-th component of  ${\bf r}_{\zw}$, 
$r_{\zw,i}$ say, for ${\bf r} =
{\bf a},{\bf b}, {\bf c}$, equals $r_{\zw,i}^+$ when $G_i > 0$, and
$r_{\zw,i} = r_{\zw,i}^-$ for $G_i < 0$.
The necessity for left and right analyticity emerges in the case of
special geometries, i.e. as a consequence of
the actual shape of the particles and of 
the medium where they move. There are cases, indeed, such that
reversing the sign of the components of ${\bf G}$ may require
a change in the values of the constants ${\bf a}$, ${\bf b}$ and ${\bf c}$
above, so that only the left and the right analyticity around
$G_i = 0$, for all $i$, hold. 
For our purpose, this is sufficient, but in the 
case we have full analyticity at ${\bf G} = 0$,
some approximations can be improved, as explained below, after Theorem 1.

Assumptions 1. and 2. are needed here to justify the 
calculations of subsections 2.3 and 2.4 but, similarly to the assumptions
of Refs.\cite{GC1,GC2}, they cannot be validated at present
on the sole grounds of the given dynamics. Only {\it a posteriori} we can 
check whether using them we are led into some kind of inconsistency or not.
However, to accept them as possibly valid ({\it a priori}) we 
may rely on the evidence accumulated so far in the literature, e.g. on
several studies of the nonequilibrium Lorentz gas like 
Refs.\cite{V,LNRM,CELS} and others.
In particular, Assumption 2.a is also in agreement with the discussion
of pp.949-950 of Ref.\cite{GC2}, where it is argued that the transition
from equilibrium to nonequilibrium stationary states, with small
forcing, should be seen as producing ``... an insignificant deformation
of the unstable manifold $W_O^u$ ...'', i.e. of the unstable manifold
of a given periodic point. Similarly, one may argue that the
properties of periodic orbits will only change little for small
changes in the fields. As a matter of fact,
the periodic (trianglar lattice) Lorentz gas has been used to test
many orbits 
to high numerical precision in order to see whether their 
contributions to numerator and denominator of Eq.(\ref{POEJ}) are 
smooth in the field. 
The result is that they appear to be smooth \cite{LNRM}, 
with the only exceptions
of orbits with grazing collisions, i.e.
orbits which suddenly appear or disappear from the phase space,
when the field is varied ever so slightly. However, it does not seem  
that such orbits should concern us. In the first place,
they do not exist if the interactions are soft, and hard core
interactions could be seen as a limiting case of soft core ones,
Ref.\cite{BEC}.
Also, the numerical tests of
Refs.\cite{MRC,MR,LNRM} have not evidenced any difficulty 
linked to this problem,
because the orbits with grazing collisions were a very small
fraction of the whole, with negligible weight each. 
In the Appendix, we give an argument
to justify why these orbits
should always be assigned a vanishing weight in the expansion 
Eq.(\ref{POEJ}). 

\noindent
A few further remarks are in order.

\newtheorem{remark}{Remark}

Assumption 2. plays for us the role of the differentiability of SRB 
measures in Ref.\cite{GR}, but it may appear as rather strong.
On the other hand,
this assumption may be relaxed in various ways, and
it is supported to a good extent
both by numerical studies of simple systems, and by intuition, as
noted above. In practice, we observed that for small changes in the 
field the orbits change very little in shape which,
in turn, determines
their properties such as their period, stability, $\za$ etc.
Furthermore, it is important to note that we do not require that our 
conditions
hold for very general functions of phase. On the contrary,
we restrict the validity
of our conditions to very special quantities $-$the phase space 
contraction rate and its derivatives$-$ which are easily seen to
have particularly good 
properties in the dynamics of many interesting maps.

A comparison between Assumption 1. and the assumptions of Ref.\cite{GC2}
shows that the latter directly refer to the dynamics of the system,
i.e. to the properties of the equations of motion and of the space
in which the motion takes place. Our assumption, instead, refers to
the stationary measure, which is not the dynamics but only a result 
of the dynamics. Therefore, Assumption 1. is valid if the system is Anosov 
or axiom-A (hence, also if the assumptions of Ref.\cite{GC2} hold)
but it could still remain valid even for dynamics
of a different kind. On the other hand, our assumption is not as
fundamental because, for instance, it does not provide us with a
direct way of estimating the errors connected with the $n$-th order
UPO approximations of the POE.

\subsection{Derivation of the Onsager reciprocity relations}

We can now prove the validity of Onsager's
relations for the systems verifying our assumptions (which we take to
be general $N$-particle systems subject to an iso-energetic 
Gaussian thermostat). The proof proceeds as follows.
First, we consider the $n$-th order UPO approximation of 
Eq.(\ref{POEJ}), and we expand to first order in ${\bf G}$ the 
resulting expressions.
Then, we group together the contributions of periodic orbits
with opposite currents, in order to use the relation called 
{\it Lyapunov sum rule} in Ref.\cite{RMlarge}. This is a relation
between the parameter $\za$ and the Lyapunov exponents associated 
with a given periodic orbit (see Eq.(\ref{srule}) below). This allows us
to find a symmetry between different derivatives of the 
$n$-th order UPO approximate current. Finally, exploiting
Assumption 2. we interchange the derivative with
taking the infinite period limit operation, to obtain the desired result. 
The major ingredient in these calculations
is the time reversal symmetry of the equations of motion,
together with the density of the attractor in the phase 
space. Indeed, it is because of these two features that we can write 
Eqs.(\ref{rtX},\ref{grouped},\ref{determ}) below.

Using the assumed analyticity in the fields
of the period and of the stability weight (Eqs.(\ref{tayT},\ref{tayL})),  
we can expand the denominator of Eq.(\ref{Jk}) to first order in
${\bf G}$ and obtain
\newpage
\begin{eqnarray}
& & \langle {\bf J} \rangle^{(n)}({\bf G}) = \nonumber \\
& & \quad \frac{\sum_{\zw \in P_n(G)} \Lambda_{\zw,u}^{-1} {\bf I}_{\zw}}
{\sum_{\zw \in P_n(G)} \zt_{\zw(0)} \zL_{\zw(0),u}^{-1}} \left[
1 - \frac{\sum_{\zw \in P_n(G)} \zt_{\zw(0)} {\bf b}_{\zw} 
+ \zL_{\zw(0),u}^{-1}
{\bf a}_{\zw}}{\sum_{\zw \in P_n(G)} \zt_{\zw(0)} \zL_{\zw(0),u}^{-1}} \cdot 
{\bf G} 
+ o(G^2) \right] ~. \quad \quad 
\label{firstexp}
\end{eqnarray}
Then, the time
reversibility of Eqs.(\ref{eqsmot}), combined with the density of
the relevant attractor in phase space,  guarantee that
for every orbit $\zw \in P_n(G)$ there is another orbit
$-\zw \in P_n(G)$ with equal period and opposite cumulative 
current ${\bf I}$,\footnote{Observe that
time reversal symmetry in the equations of motion is not enough for our 
purposes (see e.g. Ref.\cite{BG}). 
The equations of motion of our models are time reversible for
all {\bf G}, however 
the corresponding attractors and repellers may cover disjoint regions of
phase space if {\bf G} is sufficiently large. In that case, 
taking the time reverse image of a point in the
attractor produces a point in the repeller, and it is not true 
then that $-\zw \in P_n(G)$ if $\zw \in P_n(G)$. 
On the contrary, the time reverse of a UPO in $P_n(G)$
is still in $P_n(G)$, if the attractor is dense (as in Anosov systems). 
Systems with non dense attractors can be studied (see e.g. Ref\cite{BG}), 
but they do not concern us here, as we are interested in the small 
fields regime.} i.e.:
\begin{equation}
\tau_{-\zw(G)} = \tau_{\zw(G)} ~, \quad
{\bf I}_{-\zw(G)} = - {\bf I}_{\zw(G)} ~. 
\label{rtX}
\end{equation}
Hence, by grouping together such pairs of 
orbits, we get:
\begin{equation}
\sum_{\zw \in P_n(G)} \Lambda_{\zw,u}^{-1} {\bf I}_{\zw} = 
\sum_{\zw \in P_n^+(G)} \left( \Lambda_{\zw,u}^{-1} -
\Lambda_{-\zw,u}^{-1} \right) {\bf I}_{\zw} \nonumber 
= \sum_{\zw \in P_n^+(G)} {\bf I}_{\zw} \Lambda_{\zw,u}^{-1} 
\left( 1 - \Lambda_\zw \right) ~,
\label{grouped}
\end{equation}
where $\zw \in P_n^+(G)$ if it has
$n$ collisions and $\int_\zw \za(S_t \zG_\zw) d t > 0$,\footnote{Given
$n \in \zN$, it may happen that there 
are no UPOs in $P_n(G)$ with $\int_\zw \za(S_t \zG_\zw) d t > 0$, 
in which case $P_n^+(G)$ is empty and the sums in Eqs.(\ref{grouped})
vanish. If this happens for all $n$ greater or equal to
a given $n_0$, then 
$\langle {\bf J} \rangle_\zW =0$, and our calculations 
leading to Eq.(\ref{OReq}) trivially hold.}
and
\begin{equation}
\Lambda_{\zw} \equiv \exp \left( \tau_{\zw} 
\sum_{j=1}^{2dN} \lambda_{\zw,j} \right) \, ,
\label{determ}
\end{equation}
where the sum this time involves {\it all} 
the Lyapunov exponents of $\zw$.
The Lyapunov sum rule for a periodic orbit $\zw$
-Eq.(8) in Ref.\cite{RMlarge}- in our context can be written as
\begin{eqnarray}
\zt_\zw \sum_{j=1}^{2dN} \zl_{\zw(G),j} &=& -(dN-1) \int_{\zw(G)}
\za(S_t \zG_{\zw(G)}) ~d t \nonumber \\
&=& - (dN-1) \int_{\zw(G)} {\bf c}_{\zw(0)}(S_t \zG_{\zw(0)}) ~ dt 
\cdot {\bf G} + o(G^2) ~.
\label{srule}
\end{eqnarray}
Note that for ${\bf G}=0$
the flow does not contract or expand elements of phase space, i.e.
$\zL_{\zw(G=0)} = 1$ for all UPOs, and each addend in the sums in 
Eq.(\ref{grouped}) vanish. Then, 
expanding the terms in Eq.(\ref{grouped}) in powers of ${\bf G}$, 
and substituting into Eq.(\ref{firstexp}) we obtain
\begin{equation}
\langle {\bf J} \rangle^{(n)}({\bf G}) = {\bf A}^{(n)} {\bf G} + o(G^2) ~,
\label{secexp}
\end{equation}
where ${\bf A}^{(n)}$ is the symmetric matrix whose $ij$-entry is
\begin{equation}
A_{ij}^{(n)} = (dN - 1)
\frac{\sum_{\zw \in P_n(G)} \zL_{\zw(0),u}^{-1} Q_{\zw(0),i} Q_{\zw(0),j}}
{\sum_{\zw \in P_n(G)} \zt_{\zw(0)} 
\zL_{\zw(0),u}^{-1}} ~,
\label{approx}
\end{equation}
and $Q_{\zw(0),i}$ is defined by
\begin{equation}
Q_{\zw(0),i} = \int_{\zw(G)} c_{\zw(0),i}(S_t \zG_{\zw(0)}) ~ dt ~.
\label{Qterm}
\end{equation}
Thus, if we differentiate the $k$-th component of the approximate
current, $\langle J_k \rangle^{(n)}({\bf G})$, 
with respect to $G_l$ say, we obtain
\begin{equation}
\partial_{G_l} \langle J_k \rangle^{(n)}({\bf G})  =
A_{kl}^{(n)} + o_1(G) = A_{lk}^{(n)} + o_2(G) =  
\partial_{G_k} \langle J_l \rangle^{(n)}({\bf G}) + o_3(G) ~,
\end{equation}
where the $o_i(G)$, $i=1,2,3$, are distinct quantities of order $G$.
Let us denote by 
\begin{equation}
L_{kl} \equiv \left. \partial_{G_l} \langle J_k \rangle_{_{\zW}}
\right|_{{\bf G}=0} = \left. \partial_{G_l} \lim_{n\rightarrow\infty}
\langle J_k \rangle^{(n)} \right|_{{\bf G}=0} ~,
\end{equation}
the $kl$-entry in the transport coefficients tensor, and recall that
the sequence $\{ \langle J_k \rangle^{(n)} ({\bf G}) \}_{n=1}^\infty$ 
converges to a limit (Assumption 1) uniformly (Assumption 2),
as $n \rightarrow \infty$. Then, we have (Assumptions 1. and 2.)
\begin{equation}
L_{kl} = \left. \partial_{G_l}
\lim_{n \rightarrow \infty} \langle J_k \rangle^{(n)} 
\right|_{{\bf G}=0} 
= \lim_{n \rightarrow \infty} \left.
\partial_{G_l} \langle J_k \rangle^{(n)} \right|_{{\bf G}=0} =
L_{lk} ~; \quad k,l = 1, ..., \zn ~.
\label{OReq}
\end{equation}
These are the Onsager relations for our systems, and we can state that:
\newtheorem{proposition}{Theorem} 
\begin{proposition}
For thermostatted $N$ particle systems verifying Assumptions 1. and 2. 
the following holds
\begin{equation}
L_{kl} = L_{lk} \quad \quad \mbox{for all} \quad k,l = 1, ..., d ~.
\end{equation}
\end{proposition}

As mentioned above, we note that the geometry at hand may improve 
our calculations, thanks to ensuing symmetries of the
corresponding systems. 
By geometry we mean all the specifications
which have to do with the shape of the particles and of the 
medium in which they move. In most cases, these are such that 
reversing the direction of ${\bf G}$ effectively amounts to just a 
rotation of the coordinate axes. In other words, the UPOs embedded
in the attractor corresponding to $-{\bf G}$ have the same properties 
as those in the attractor corresponding to ${\bf G}$, except 
for having an opposite cumulative current: i.e. for every $n \in \zN$, 
and for every
$\zw(G) \in P_n(G)$ there is $\zw(-G) \in P_n(-G)$ such that
\begin{equation}
\zt_{\zw(-G)} = \zt_{\zw(G)} ~; \quad 
\zL_{\zw(-G),u}^{-1} = \zL_{\zw(G),u}^{-1} ~; \quad
{\bf I}_{\zw(-G)} = - {\bf I}_{\zw(G)} ~. \label{TR3}
\end{equation}
Observe that this has nothing
to do with time reversal invariance, although it may appear to be 
similar. Equations (\ref{TR3}) are a consequence of the geometry of the 
system only, and express what can be referred to as
the $[{\bf G} \rightarrow -{\bf G}]$-field reversal symmetry,
which may not obtain for anisotropic systems. Therefore,
having distinguished the time reversal invariance from the 
(possibly not present) field reversal symmetry, we see that the second 
is not needed for the validity of the Onsager reciprocity relations.
If, on the other hand, Eqs.(\ref{TR3}) hold, Assumption 2. can be taken in 
the sense of full analyticity,\footnote{For instance,
in Ref.\cite{G2} full analyticity of $\za$ around ${\bf G} = 0$ is assumed.} 
and the terms
${\bf I}_{\zw(G)}(\zL_{\zw(G),u}^{-1} - \zL_{-\zw(G),u}^{-1})$
of Eq.(\ref{grouped}) are odd in ${\bf G}$, while the 
terms $\zt_{\zw(G)} \zL_{\zw(G),u}^{-1}$ of Eq.(\ref{Jk})
are even in ${\bf G}$. Then,
Eq.(\ref{secexp}) can be re-written as
\begin{equation}
\langle {\bf J} \rangle^{(n)}({\bf G}) = {\bf A}^{(n)} {\bf G} +
o(G^3) 
\end{equation}
making our calculations correct to second order, rather than just
first order in the field. However, such field 
reversal symmetry is not needed for the derivation of the OR.

\subsection{Nonnegativity of transport coefficients}

Here, we generalize an argument recently proposed in Ref.\cite{RMlarge}.
Let us consider the transport coefficient associated with the $k$-th
current:
\begin{equation}
\zs_k = \partial_{G_k} \langle J_k \rangle_\zW |_{G=0} = L_{kk} ~, 
\quad \quad k = 1, ..., \zn ~.
\end{equation}
Then, in view of Eq.(\ref{POEJ}), and 
grouping terms as in Eq.(\ref{grouped}),
allows us also to deduce the fact that $\zs_k$ cannot be 
negative. To see that,
rewrite Eq.(\ref{POEJ}) as
\begin{equation}
\langle J_k \rangle_{_{\zW}}({\bf G}) = \lim_{n \rightarrow \infty}
\frac{\sum_{\zw \in P_n^+(G)} 
\zL_{\zw,u}^{-1} I_{\zw,k} \left[ 1 -  \zL_{\zw} \right]}{
\sum_{\zw \in P_n(G)} \zt_\zw \zL_{\zw,u}^{-1}} 
= \sum_{i=1}^\zn A_{ki} G_i + o(G^2) ~; \quad k = 1, ..., \zn
\end{equation}
which implies $\zs_k = A_{kk}$. This quantity, in turn, is nonnegative 
as it results from taking the limit  $A_{kk} \equiv
\lim_{n\rightarrow\infty} A_{kk}^{(n)}$ in
Eq.(\ref{approx}) (the limit exists because of Assumptions 1. and 2.). 
Therefore, we can state this as
\begin{proposition}
For thermostatted $N$ particle systems satisfying Assumptions 1. and 2, 
$\zs_k \ge 0$, for all $k =1,...,\zn$. 
\end{proposition}

For this result Assumptions 1. and 2. are enough.
On the contrary, the problem of the
strict positivity of $\zs_k$ does depend on further conditions,
similarly to the situation investigated in Ref.\cite{RU0}.  

\vskip 15pt

We conclude this section by observing that all our results
are not purely dynamical in nature: they rest
on the choice of the initial ensemble, which must evolve into
the stationary SRB measure by the given dynamics. Indeed, 
initial conditions such that the consequent evolution violates our
conclusions are possible.
However,
Assumption 1. says that such initial conditions are very special,
because they only constitute a set of zero Lebesgue measure. 
We also observe that the identification made in Eq.(\ref{currentdef})
seems to require a large number of particles $N$, in order to
make physical sense (see Ref.\cite{CR1}).

\section{Discussion}
The purpose of this section is to put forward possible problems/open
questions in our approach, in order to stimulate
further research.

\vskip 5pt \noindent
1. First of all, we have tried to look into the results of 
Refs.\cite{G1,G2,GR} from a different perspective, and we have found
that (for the purpose of this work) the assumptions of those
papers could be replaced by others. 
Indeed, rather than saying that our systems are Anosov-like 
(or axiom A-like), we merely use one representation of the average  
currents, Eq.(\ref{POEJ}), which would be correct for all smooth
functions if the systems were 
really Anosov, and we postulate
that the currents associated with UPOs are well behaved.
This is similar to equilibrium statistical mechanics where
rather than specifying the expected chaotic properties of
the given systems,
the form of the ensembles are postulated, and calculations are developed
afterwards with such ensembles. One advantage of that might be
that we do not really have to know which properties of the dynamics
of particle systems make them ``Anosov-like''. In particular,
smoothness of the dynamics, hyperbolicity etc. may not be strictly
necessary, as several examples seem to indicate, 
especially if only selected functions of phase are considered.
For instance, the Lorentz gas
has singularities (although it is strongly hyperbolic); while the Ulam map
is not hyperbolic. Yet, POE have performed correctly when applied
to such systems (see Ref.\cite{MRaust} for the Ulam map).
See also Ref.\cite{DC}, in which intermittent 
diffusion is treated in terms of cycle expansions.

\vskip 5pt \noindent
2. Another advantage of our approach is precisely the fact that
the special representation of the stationary state which we use
$-$that based on orbital measures$-$ can actually and succesfully be
implemented in the simplest particle systems (various periodic Lorentz 
gas models), at least in the calculations
of several quantities \cite{MRC,CGS,V,MR,LNRM,DM}. For the same reason,
our predictions formulated in terms
of POE are amenable to direct tests
via computer simulations of such systems.

\vskip 5pt \noindent
3. In all approaches, ours
as wel as those of Refs.\cite{G1,G2,GR}, the necessity of having a 
large number of particles seems to be absent. Classical
proofs of the Onsager relations
always relied upon the assumption that the system
is large (made of $N \approx 10^{23}$ particles), and 
that it can be split into small ``local'' ones; basic
variables $a_1, ..., a_n$, with $n \ll N$ were associated with these
small local subsystems ($n \approx 10^{10}$ say).
The origin of this apparent paradox is the assumed equality of the 
average phase space contraction and macroscopic 
entropy production rates, which only seems to
hold for large $N$, Ref.\cite{CR1}.

\vskip 5pt \noindent
4. In the paper we have avoided the problem of dealing with the singular
points of the dynamics of systems of hard particles, saying that
they would only get vanishing weights in the POE Eq.(\ref{POEJ}).
In the Appendix, we give an argument to support this point of view.
If that is correct, and orbits with grazing collisions effectively
do not spoil the POE-representation of the stationary 
distributions, because they do not contribute to Eq.(\ref{POEJ}), 
then we have evidenced one possible mechanism through which the techniques
devised for axiom-A systems remain valid in more general settings,
e.g. those of Refs.\cite{MRC,CGS,V,MR,LNRM,DM}.

\vskip 20pt

\noindent
{\Large \bf Appendix}
\vskip 10pt 
\noindent
Here we argue that in case of hard core interactions, UPOs
with grazing collisions can be neglected in the expansion Eq.(\ref{POEJ}),
so that they do not endanger the validity of Assumptions 1. and 2.
First of all, the largest Lyapunov exponent
of one such orbit would appear to be very large
(effectively infinite) in numerical simulations; 
hence a very small (effectively vanishing) weight 
would be assigned to such orbits (see e.g. Ref.\cite{Sa}).
Indeed, consider the trajectories of two particles which are going to
experience a grazing collision.
If a perturbation moves the trajectories of the two particles
a little away 
from the colliding path, there will be no collision at all,
while there will be a non grazing collision if the perturbation moves the
trajectories closer to each other.

Moreover, for the Lorentz gas at equilibrium, these ideas 
can be supported by an analytical argument.
The Lyapunov exponents, in this case, 
are obtained from the logarithms of the eigenvalues
of products of matrices of the following form:\footnote{This
follows from a simple calculation. See, e.g., Ref.\cite{PR}.}
\begin{equation}
{\cal M} \equiv \prod_{j=1}^n {\cal M}_j ~,
\end{equation}
if $n$ is the number of collisions, where
\begin{equation}
{\cal M}_{j} = - \left( \begin{array}{cl}
1 & \frac{2}{\cos \psi_j} \\
l_j & 1 + \frac{2 l_j}{\cos \psi_j} \end{array} \right) ~.
\label{collisj}
\end{equation}
Here, $l_j$ is the distance
travelled by the moving particle between the collisions $j$ and $j+1$,
and $\psi_j$ is the collision angle at the $j$th collision, where
$\psi_j = 0$ means head on collision, while $\psi_j = \pi/2$
corresponds to a grazing collision.
Because $\psi_j \in [0,\pi/2]$, and $\cos \psi_{j} \ge 0$ for all $j$,
every ${\cal M}_j$ is defined
and consists of negative entries if $\psi_j \in [0,\pi/2)$ (hard collisions),
while it is not defined for grazing collisions.  

Let us compute ${\cal M}$ for an orbit with $n$ collisions,
and $\psi_{j} < \pi/2$ for all $j = 1, ..., n$, and denote by
\begin{equation}
{\cal M}^{(n-1)} = \prod_{j=1}^{n-1} {\cal M}_j \equiv (-1)^{n-1} 
\left( \begin{array}{cc}
a & b \\
c & d \end{array} \right) ~,
\end{equation}
the product of the first $(n-1)$ blocks.
The terms $a,b,c$ and $d$ are then finite,
positive numbers. Now, multiply ${\cal M}^{(n-1)}$ by ${\cal M}_n$,
the matrix of the last free flight and collision,
in order to obtain ${\cal M}$:
\begin{eqnarray}
{\cal M} &=& {\cal M}_n {\cal M}^{(n-1)} = (-1)^n  
\left( \begin{array}{cc}
1 & 2/\cos \psi_n \\
l_n & 1 + 2 l_n/\cos \psi_n \end{array} \right)
\left( \begin{array}{cc}
a & b \\
c & d \end{array} \right) \\ 
&=& (-1)^n \left( \begin{array}{cc}
a + \frac{2 c}{\cos \psi_n} & b +\frac{2 d}{\cos \psi_n} \\
a l_n + c \left(1 + \frac{2 l_n}{\cos \psi_n} \right) & 
b l_n + d \left(1 + \frac{2 l_n}{\cos \psi_n} \right) \end{array} \right)
~.
\end{eqnarray}
The eigenvalues of ${\cal M}$ obey
\begin{equation}
\zL_u + \zL_s = \mbox{Tr}({\cal M}) = (-1)^n 
\left( a + d + b l_n + \frac{2}{\cos \psi_n}
(c + d l_n) \right) ~, 
\end{equation}
and 
\begin{equation}
\zL_u \zL_s = \mbox{det} ({\cal M}) = \pm 1 ~.
\end{equation}
Hence, the determinant remains bounded, while
the magnitude of the trace grows without limits 
if $\psi_n \rightarrow \pi/2$, which implies that
one of the eigenvalues becomes very large, while the other
becomes very small. The result is that the 
weight $\zL_u^{-1}$ in Eq.(\ref{POEJ}) gets smaller and 
smaller, and converges to zero for UPOs with $\psi_n$
closer and closer to $\pi/2$. 

This argument is not yet a full proof
that orbits with grazing collisions can be neglected in the POE,
since we have not explained in which sense one such orbit could 
be considered as the limit of a sequence of orbits without 
grazing collisions.
However, the argument shows how a collision which is sufficiently
close to grazing contributes to make the weight of the 
corresponding orbit small and, keeping fixed the other quantities,
the weight is the smaller, the closer to
grazing the collision is.
We conclude noting that the problem posed by 
trajectories with grazing collisions is present in all approaches
based on dynamical weights.
In this respect, one of the nice features of the approach based
on periodic orbits is that each UPO,
as a whole, covers only a {\em finite} length, hence the relevant
stable and unstable manifolds, and the UPO's contribution
to the POE can be computed in finitely many steps. This is why 
the previous analysis for the Lorentz gas can be carried out.

\vskip 30pt

\noindent
{\Large \bf Acknowledgements} 
\vskip 10pt 
\noindent
This work has been supported in part by GNFM-CNR (Italy), and
through the grant ``EC contract ERBCHRXCT940460" for the project
``Stability and universality in classical mechanics".
In particular GNFM-CNR supported one of the author's (EGDC)
stay in Torino, where the ideas originated.
EGDC also gratefully acknowledges support from the US Department of Energy
under grant DE-FG02-88-ER13847.

\end{document}